\documentstyle[12pt]{ioplppt}          
\def\cqo#1#2{\cos[q\Omega^{\rm #1}_S(b_f,R^s_{#2})]}
\def\ut#1{\mathop{\vtop{\ialign{##\crcr
     $\hfil\displaystyle{#1}\hfil$\crcr\noalign
     {\kern1pt\nointerlineskip}\hbox{$\hfil\sim\hfil$}\crcr
     \noalign{\kern1pt}}}}}

\def\undersymbol#1#2{\mathop{\vtop{\ialign{##\crcr
     $\hfil\displaystyle{#2}\hfil$\crcr\noalign
     {\kern1pt\nointerlineskip}\hbox{$\hfil#1\hfil$}\crcr
     \noalign{\kern1pt}}}}}

\def\degr{^0}

\begin{document}
\title{Gamma Ray Emission From A Baryonic Dark Halo~\footnote{We would like 
to dedicate this work to the memory of Dennis W. Sciama}}

\author{F De Paolis\dag, G Ingrosso\dag, Ph Jetzer\ddag\ftnote{3}{To
whom correspondence should be addressed.} and 
M Roncadelli\P}
\address{\dag\ Dipartimento di Fisica, Universit\`a di Lecce and INFN, 
Sezione di Lecce,  CP 193, I-73100 Lecce, Italy}
\address{\ddag\ Paul Scherrer Institute, Laboratory for Astrophysics, 
CH-5232 Villigen PSI, and Institute of Theoretical Physics, 
University of Zurich, Winterthurerstrasse
190, CH-8057 Zurich, Switzerland}
\address{\P\ INFN, Sezione di Pavia, Via Bassi 6, I-27100, Pavia, Italy}

\begin{abstract}
A re-analysis of EGRET data by 
Dixon et al. \cite{dixon} has led to the 
discovery of a statistically
significant diffuse $\gamma$-ray emission from the galactic halo. 
We show that this emission can naturally be accounted for within a 
previously-proposed model for baryonic dark matter, 
according to which dark clusters
of brown dwarfs and cold self-gravitating $H_2$ clouds populate the outer 
galactic halo and can show up in microlensing observations.
Basically, cosmic-ray protons in the galactic halo scatter on the clouds 
clumped into dark clusters, giving rise to the observed $\gamma$-ray flux. 
We derive maps for the corresponding intensity distribution, which turn
out to be in remarkably good agreement with those obtained by Dixon et al. 
\cite{dixon}. 
We also address future prospects to test our predictions.
\end{abstract}


\section{Introduction}
Observations of the diffuse $\gamma$-ray emission during the last 
twenty years \footnote{A comprehensive account of these matters as well as 
of their theoretical explanations can be found in \cite{ft}.} 
have been successfully interpreted in terms of a two-folded structure \\
$\star$ a highly anisotropic component strongly concentrated along the 
galactic disk,  \\
$\star$ an apparently isotropic component. \\
While the former is evidently galactic in nature - being actually
accounted for by cosmic ray (CR) interactions in the interstellar medium (ISM) 
\cite{hunter} -
the origin of the latter still remains an open problem
in high-energy  astrophysics (see e.g. \cite{dixon,pohl,smr}).
We will restrict our attention to the latter component throughout the 
present paper.

We begin by recalling that EGRET observations have 
detected a diffuse $\gamma$-ray flux \cite{sreekumar}
\begin{equation}
\Phi_{\gamma}(E_{\gamma}>0.1 {\rm GeV}) = (1.45 \pm 0.05)  \times
10^{-5}~\gamma~{\rm cm^{-2}~s^{-1}~sr^{-1}}~, \label{2}
\end{equation}
with a spectral slope of $-2.10 \pm 0.03$, which -
for $E_{\gamma}>1$ GeV - gives
\begin{equation}
\Phi_{\gamma}(E_{\gamma}>1 {\rm GeV}) = (1.14 \pm 0.04)  \times
10^{-6}~\gamma~{\rm cm^{-2}~s^{-1}~sr^{-1}}~.
\label{eq:3}
\end{equation}

A question naturally arises. Where does the $\gamma$-ray emission in 
question come from? No doubt, its characteristic isotropy  calls for an 
extragalactic origin - an option which is further supported by the fact 
that it fits remarkably well with the extragalactic hard X-ray background 
\cite{thompson}. 

The next question to address is whether the considered $\gamma$-ray 
background arises from a truly diffuse process or rather from the 
contribution of very many unresolved point sources. Either option has 
received considerable attention. Among the theories of diffuse origin are a 
baryon-symmetric Universe \cite{smb}, 
primordial black hole evaporation 
\cite{ph,cline}, early 
collapse of supermassive black holes \cite{go}, a new 
population of Geminga-like pulsars \cite{hart} and WIMP (Weakly 
Interacting Massive Particle) annihilation 
(see e.g. \cite{jkg}).
Models based on discrete source contribution include  a variety of 
possibilities. What is clear since  a long time is that normal galaxies 
fail to account for the observed isotropic background -
at least as long as their disk emission is considered
\cite{fichtel}-\cite{gal3} - since the 
corresponding intensity falls shorter by a factor $\sim 10$ with respect to 
the detected flux. 
A more realistic option is provided by active galaxies 
\cite{active1,active2}.
Indeed, blazars seem to yield a successful explanation of the isotropic 
$\gamma$-ray emission \cite{blazars1}-\cite{blazars5}.
Finally, a somewhat hybrid model has recently been proposed, in which the 
isotropic $\gamma$-ray background is produced in clusters of galaxies 
through the interaction of CRs with the hot intracluster gas \cite{dar}. 
However, this model has been severely criticized 
\cite{ss,bbp}.
In fact, it gives rise to a $\gamma$-ray spectral index in disagreement 
with the observed one 
and relies upon a value for the CR density in the 
intracluster space which is too high to be plausible.
More generally, it has been shown that the contribution to the 
isotropic $\gamma$-ray emission from clusters of galaxies is  
negligible \cite{bbp}.

Recently, Dixon et al. \cite{dixon} 
have re-analyzed the EGRET data concerning the diffuse $\gamma$-ray flux 
with a wavelet-based technique,
using the expected (galactic plus isotropic) emission as a null hypothesis.
Although the wavelet approach does not allow for a good estimate of the errors,
they find a statistically significant diffuse emission from 
an extended  halo surrounding the Milky Way. 
This emission 
traces a somewhat flattened halo and its 
intensity at high-galactic latitude is \cite{dixon}
\begin{equation}
\Phi_{\gamma}(E_{\gamma}>1 {\rm GeV}) \simeq 10^{-7}-10^{-6}
~\gamma~{\rm cm^{-2}~s^{-1}~sr^{-1}}~. 
\label{eq:4}
\end{equation}
Clearly, the comparison of eqs. (\ref{eq:3}) and (\ref{eq:4}) entails
that the newly discovered halo $\gamma$-ray flux is a relevant fraction
of the standard isotropic diffuse emission  (at least for $E_{\gamma} >1$ GeV).

Our aim is to show that the observed halo 
$\gamma$-ray emission naturally arises 
within a previously-proposed model for baryonic dark matter, 
according to which dark clusters of brown dwarfs and cold 
self-gravitating $H_2$ clouds populate the outer 
galactic halo and can show up in microlensing observations 
\cite{depaolis1}-\cite{depaolisapj}.
Basically, CR protons in the galactic halo scatter on the clouds 
clumped into dark clusters, 
giving rise to the newly discovered $\gamma$-ray flux.

Although we already pointed out that a signature of the model is a diffuse 
$\gamma$-ray emission from the galactic halo \cite{depaolis1,depaolis2}, a 
more thorough  study  is 
required to compare the predicted intensity 
distribution with the observed one.  A short account of these results has 
been presented elsewhere \cite{depaolisapjl}. In the present paper, we provide 
a more exhaustive analysis. In addition,
we estimate the $\gamma$-ray emission from the nearby M31 
galaxy.


The paper is organized as follows. 
In Section 2 we recall the main 
features of our model for baryonic dark matter in the galactic halo.
In Section 3 we address the CR confinement
in the galactic halo and we estimate the CR energy density. 
In Section 4 we compute the halo
$\gamma$-ray flux - produced by the clouds
clumped into dark clusters through proton-proton scattering -
as detected on Earth.
Section 5 is devoted to the study of the $\gamma$-ray flux due to
Inverse Compton (IC) scattering of electrons off background  photons.
In Section 6 we present $\gamma$-ray intensity maps, pertaining to both 
proton-proton scattering and IC scattering, and discuss their interplay.
Finally, in Section 7 we address future prospects to test our predictions.

\section{Dark clusters in the galactic halo}

Ever since the discovery that standard big-bang nucleosynthesis correctly 
accounts for the light element abundances, a lesson has become clear: most of 
the baryons in the Universe happen to be in nonluminous form, thereby 
making a strong case for baryonic dark matter. 

In order to see how this comes 
about, we recall that the fraction of critical density contributed by luminous 
matter is estimated to be $\Omega_L \sim 0.005$~\cite{bld} 
\footnote{We are using throughout the presently favoured value 
of the Hubble constant 
$H_0 \simeq 70$ km s$^{-1}$ Mpc$^{-1}$.}. 
Yet, agreement between the predicted and observed abundances 
of nucleosynthetic yields is achieved only provided the similar 
contribution from baryons  - in whatever form - lies in the 
range $0.01  \ut <\Omega_B  \ut <0.05$~\cite{Schr98}. Actually, this conclusion has recently been sharpened 
by deuterium measurements in Quasi Stellar Object (QSO) absorption 
spectra, which probe regions of 
space much farther away than previously explored and  give $\Omega_B \simeq 
0.05$~\cite{Tyt96}. So, about $90\%$ of the baryonic matter in the Universe is 
expected to be dark. 

Needless to say,  one is naturally led to wonder about the 
distribution and form of baryonic dark matter. 

Several possibilities have been contemplated over the last few years. Although 
no logically compelling reason in favour of any particular option has emerged 
so far, it looks intriguing that a naturalness argument  
strongly suggests that the galactic dark halos should be 
predominantly baryonic. 

Basically, the idea is as 
follows. As is well known, both optical and HI observations have shown that all 
galactic rotation curves exhibit a universal qualitative behaviour: after a 
steep rise corresponding to the bulge, they stay approximately constant out to 
the last measured point. This feature -- namely the lack of a keplerian 
fall-off -- provides a stark evidence in favour of a spheroidal dark halo 
surrounding the luminous part of any galaxy. This is however not the end of the 
story. For, rotation curves trace the luminous -- hence baryonic -- matter 
within the optical disk, but are dominated by the halo dark matter at larger 
galactocentric distances. Yet, both contributions invariably turn out to match 
smoothly and exactly, thereby signalling a striking visible-invisible 
conspiracy (also called disk-halo conspiracy). Before proceeding further, a 
point should be stressed. With only a rather limited sample of available 
rotation curves, that conspiracy was initially understood as a fine-tuning 
whereby the disk and the halo of spiral galaxies manage to produce 
a flat rotation 
curve~\cite{Alb86}. Further studies have shown that such a flatness is only 
approximate: brighter galaxies tend to have slightly falling rotation curves, 
whereas fainter ones possess slightly rising rotation curves~\cite{Per91}. 
Still, what really matters for the visible-invisible conspiracy (as 
stated above) is the lack of any jump in the rotation curve within the 
disk-halo transition region, besides the approximate flatness. 

A priori, only a mysterious fine-tuning could 
justify the conspiracy in question if the halo dark matter were 
different in nature from luminous matter, that is to say if it were
nonbaryonic. So, baryonic dark matter looks 
like a  natural constituent of galactic halos. 
Incidentally, this situation is very reminiscent of the case of grand unified 
theories in particle physics, where supersymmetry has been invoked as a 
successful way out of a similar, mysterious fine-tuning needed to stabilize the 
gauge hierarchy against radiative corrections~\cite{Maiani79}. Thus, we are led 
to the conclusion that -- much in the same way as fundamental interactions 
ought to be supersymmetric -- galactic halos ought to be predominantly 
baryonic!

Remarkably enough, a specific model of baryonic dark halos emerges naturally  
from the present-day understanding of globular clusters. Indeed, a few years 
ago we have realized~\cite{depaolis1,depaolis2} that the Fall-Rees 
theory for the formation of globular clusters~\cite{fall}-\cite{vietri} 
automatically 
predicts -- without any further physical assumption -- that dark clusters made 
of brown dwarfs~\footnote{Although we concentrate our attention on brown 
dwarfs, it should be mentioned that red dwarfs as well can be accomodated 
within the considered setting.} and cold $H_2$ clouds should lurk in 
the galactic halo at 
galactocentric distances larger than $10-20$ kpc. Accordingly, the inner halo 
is populated by globular clusters, whereas the outer halo chiefly consists 
of dark clusters. \footnote{Similar ideas have been proposed by Ashman 
and Carr~\cite{ac}, Ashman~\cite{ashman}, Fabian and Nulsen~\cite{fn1,fn2}, and 
Kerins~\cite{kerins1,kerins2}. Moreover, a scenario almost identical to the one 
investigated here has been put forward by Gerhard and Silk~\cite{gs}. 
Somewhat different baryonic pictures have been worked out by
Pfenniger, Combes and Martinet \cite{pcm}, Sciama \cite{sciama}, and
Gibson and Schild \cite{gibson} (see also \cite{wwt}).}
Below, we summarize the main features of our model.

Although the mechanism of galaxy formation is not yet fully understood, the 
theory for the origin of globular clusters seems to be fairly well established
- thanks to the pioneering work of Fall and Rees~\cite{fall} - and can be 
summarized as follows. After its initial collapse, the proto-galaxy is expected 
to be shock heated up to its virial temperature $\sim 10^6$ K. Because of 
thermal instability, density enhancements rapidly grow as the gas cools. 
Actually, overdense regions cool more rapidly than average, and so 
proto-globular-cluster (PGC) clouds form in pressure equilibrium with the hot 
diffuse gas. When the PGC cloud temperature drops to $\sim 10^4$ K, hydrogen 
recombination occurs: at this stage, the PGC cloud mass and size are 
$\sim 10^5 (R/{\rm kpc})^{1/2} ~M_{\odot}$ and $\sim 10 (R/{\rm kpc})^{1/2}$ 
pc, respectively 
($R$ being the galactocentric distance). Below $\sim 10^4$ K, an efficient 
cooling can be brought about only by photon emission from roto-vibrational 
transitions in $H_2$. Whether this mechanism is actually operative or not 
crucially depends on the intensity of the environmental ultraviolet (UV) 
radiation field, as we are now going to discuss. 

In fact, in the central region of the proto-galaxy an AGN (Active Galactic 
Nucleus) 
along with a first population of massive stars are expected to form, which act 
as strong sources of UV radiation that dissociates the $H_2$ 
molecules. It is not difficult to estimate that the $H_2$ destruction  
should occur 
for galactocentric distances smaller than $10-20$ kpc. As a consequence, 
cooling is heavily suppressed in the inner halo, and so here the PGC clouds 
remain for a long time in quasi-hydrostatic equilibrium at temperature $\sim 
10^4$ K, resulting in the imprinting of a characteristic mass $\sim 10^6 
M_{\odot}$. Eventually, the UV flux decreases, thereby allowing for 
the formation and survival of $H_2$. Accordingly, the PGC clouds can further 
cool, collapse and fragment, ultimately producing ordinary stars clumped into 
globular clusters. 

What is most relevant for the present considerations is that in the outer halo 
-- namely for galactocentric distances larger than $10-20$ kpc -- no 
substantial $H_2$ destruction should take place, owing to the distance 
suppression of the UV flux. Therefore, here the PGC clouds 
monotonically cool, collapse and fragment. When their number density exceeds 
$\sim 10^8$ cm$^{-3}$, virtually all hydrogen gets converted to molecular form 
by three-body reactions ($H + H + H \to H_2 +H$ and $H + H + H_2 \to H_2 + 
H_2$), which makes in turn the cooling efficiency increase 
dramatically~\cite{palla}. As a result, no imprinting of a characteristic mass 
on the PGC clouds shows up, and the fragment Jeans mass can drop to values 
considerably smaller than $\sim 1 M_{\odot}$. The fragmentation process stops 
when the PGC clouds become optically thick to their own line emission -- this 
happens for a fragment Jeans mass as low as $\sim 10^{-2} 
M_{\odot}$  ~\cite{palla}. 
In this manner, dark clusters containing brown dwarfs 
in the mass range $10^{-2} - 10^{-1}~M_{\odot}$ should form in the outer 
halo. Typical values of the dark cluster radius are $\sim 10$ pc. 

In spite of the fact that the dark clusters resemble in many respects  
globular clusters, an important difference exists. Since practically no nuclear 
reactions occur in the brown dwarfs, strong stellar winds are presently 
lacking. Therefore the leftover gas - which is ordinarily expected to exceed 
60\% of the original amount - is not expelled from the dark clusters but 
remains confined inside them. Thus, also cold gas clouds are clumped into the 
dark clusters. Although these clouds are primarily made of $H_2$, they 
should be surrounded by an atomic layer and a photo-ionized ``skin''. Typical 
values of the cloud radius are $\sim 10^{-5}$ pc.

Besides accounting for the halo dark matter in a natural fashion - without 
demanding any new physical assumption - this model 
elegantly explains the visible-invisible conspiracy. For, whether ordinary  matter 
is luminous or dark ultimately depends on the intensity of the environmental UV 
radiation field during the proto-galactic epoch - no fine-tuning is indeed
involved! 
Moreover, the UV field in question is expected to be stronger for brighter 
galaxies. Accordingly, brighter galaxies should have the dark clusters 
lying farther away from the galactic centre than  fainter galaxies, 
thereby making the 
contribution of dark matter to the rotation curve of brighter galaxies less 
significant than for fainter ones: this circumstance precisely agrees with 
the above-mentioned 
observed pattern of rotation curves~\cite {Per91}.

Observationally, the present model makes a crucial prediction: very high-energy 
cosmic ray proton scattering on the clouds should give rise to a detectable 
diffuse gamma-ray flux from the halo of our galaxy. This topic will be dealt 
with in great detail in the next Sections. 

Further support in favour of the baryonic scenario 
in question comes from the understanding of the Extreme Scattering Events: 
dramatic flux changes over several weeks during monitoring of compact radio 
quasars~\cite{fiedler}. It is generally agreed that ESEs are not intrinsic 
variations, but rather apparent flux changes caused by refraction when a 
(partially) ionized cloud crosses the line of sight. Recently, 
Walker and Wardle~\cite{ww} pointed out that the first consistent 
explanation of ESEs requires the refracting clouds to have precisely the 
same properties of the 
cold $H_2$ clouds predicted by the present model 
(it is their photo-ionized ``skin'' that causes the radio wave refraction).

Last but not least is the issue of MACHOs (Massive Astrophysical Compact Halo 
Objects), detected since 1993 in microlensing experiments towards the 
Magellanic Clouds. 
Regretfully, their origin remains controversial. Although the events 
detected towards the SMC (Small Magellanic Cloud) seem to be a self-lensing 
phenomenon \cite{st, gyuk}, a similar interpretation of all the events 
discovered towards the LMC (Large Magellanic Cloud) looks unlikely 
~\cite{alcock2}.
Yet -- even if most of the MACHOs are dark matter candidates lying in the 
galactic halo -- their physical nature is unclear, since their average mass 
strongly depends on the still uncertain galactic model, ranging from 
$\sim 0.1~M_{\odot}$ for a maximal disk up to $\sim 0.5~M_{\odot}$ for a 
standard isothermal sphere. 

Superficially, white dwarfs look as the best explanation, 
but the resulting excessive metallicity of the halo makes this option 
untenable, unless their contribution to halo dark matter is not substantial
(see \cite{gm,binney}).
So, some variations on the theme of brown dwarfs have been 
explored.

An option is that the galactic halo resembles more closely a 
minimal halo (maximal disk) rather than an isothermal sphere, in which case MACHOs can 
still be brown dwarfs.~\footnote{Notice that also the $H_2$ clouds can give 
rise to microlensing events~\cite{draine}.} 
In this connection, two points should be stressed. First, a large fraction 
(up to $50\%$ in mass) can be binary systems - much like ordinary stars - 
thereby counting as twice more massive objects~\cite{depaolismnras}. 
Second, within our model brown dwarfs  can actually be beige dwarfs - with 
mass substantially larger than $\simeq 0.1~M_{\odot}$ - as suggested by 
Hansen~\cite{HANSEN}, since a slow accretion mechanism from cloud gas is 
likely to occur~\cite{lcs}.

An alternative
possibility has been pointed out by Kerins and Evans~\cite{ke}. Since in 
the present model the initial mass function obviously changes with the 
galactocentric distance,~\footnote{Evidence for a spatially varying initial 
mass function in the galactic disk has been reported~\cite{taylor}.} 
it can well happen that brown dwarfs dominate the halo 
mass density without however dominating the optical depth for microlensing. 
What are then MACHOs? Quite recently, faint blue objects discovered by the 
Hubble Space Telescope have been understood as old halo white dwarfs lying 
closer than $\sim 2$ kpc from the Sun \cite{hansen88}-\cite{ibata}: 
they look as a good 
candidate for MACHOs within this context.

Finally, we remark that recently ISO observations \cite{valentijn} of 
the nearby NGC891 galaxy have detected
a huge amount of molecular hydrogen, which
might account for almost all dark matter, at least within its optical radius. 
Other observations suggest that similar clouds are also present farther away 
\cite{lopez}.  
In addition, Sciama \cite{sciama} has argued that a known excess in the 
far-infrared emissivity of our galaxy (over that expected from a standard 
warm interstellar dust model) would be naturally accounted for by a 
population of cold $H_2$ clouds building up a thick galactic disk.


\section{Cosmic ray confinement in the galactic halo}

Neither theory nor observation allow at present to
make sharp statements about the propagation of CRs in the galactic
halo~\footnote{We stress 
that - contrary to the practice used in the CR community - 
by halo we mean the (almost) spherical galactic component which extends beyond 
$\sim$ 10 kpc.}.
Therefore, the only possibility to get some insight into this issue 
rests upon the extrapolation from the knowledge of CR propagation in the 
disk. Actually, this strategy looks sensible, since the leading effect is CR 
scattering on inhomogeneities of the magnetic field over scales
from $10^2$ pc down to less than $10^{-6}$ pc \cite{berezinskii}
and - according to our model - inhomogeneities 
of this kind are expected to be present in the halo as well,
because of the existence of molecular clouds - with a photo-ionized  
``skin'' - clumped into dark clusters. Indeed, typical values of 
the dark cluster radius are $\sim 10$ pc, whereas typical values of the 
cloud radius are $\sim 10^{-5}$ pc \cite{depaolisapj}.

As is well known, CRs up to energies of
$\sim 10^6$ GeV are confined in the galactic disk for $\sim 10^7$ yr 
\cite{berezinskii}. 
It can be shown that in the diffusion model for the propagation of
CRs, the escape time $\tau_{\it esc}$ is 
given by \cite{berezinskii}
\begin{equation}
\tau_{\it esc}\simeq\frac{R_h^2}{3D(E)}
\frac{1-
\displaystyle{ \frac{1}{2} \left(\frac{h_d}{R_h}\right)^2 } +\frac{1}{8}
\left(
\frac{h_d}{R_h}\right)^3}{1- \displaystyle{
\frac{h_d}{2R_h}}}~,
\label{tau1}
\end{equation}
where $D(E)$ is the diffusion coefficient, while $h_d$ and $R_h$ are the
half-thickness of the disk and the radius of the confinement region,
respectively. We remind that - 
for CR propagation in the disk - the diffusion coefficient is
$D(E) \simeq D_0~(E/7~ GeV)^{0.3}$ cm$^2$ s$^{-1}$ in the ultra-relativistic 
regime, whereas it reads $D(E) \simeq D_0 \simeq 3 \times 10^{28}$ cm$^2$ s$^{-1}$ in 
the non-relativistic regime \cite{berezinskii}.

CRs escaping from the disk will further diffuse in the galactic
halo, where they can be retained for a long time,
owing to the scattering on the above-mentioned
small inhomogeneities of the halo magnetic field
\footnote{
A similar idea has been proposed with a somewhat different motivation in 
\cite{wdowczyk}.}.

Indirect evidence that CRs are in fact trapped in a low-density
halo has recently been reported. For example, Simpson \&  Connell
\cite{simpson} argue that, based on measurements of isotopic abundances
of the cosmic ratio $^{26}$Al/$^{27}$Al, the CR lifetimes are perhaps a 
factor of four larger than previously thought, thereby implying that CRs 
traverse an average density smaller than that of the galactic disk. 
 
A straightforward extension of the diffusion model 
implies that the CR escape time $\tau_{\rm esc}^{~H}$ from the halo 
(of size $R_H\equiv R_h\sim 100$ kpc, much larger than the disk half-thickness) 
is given by
\begin{equation}
\tau_{\rm esc}^{~H} \simeq \frac{R_H^2}{3D_H(E)}~,
\label{tau2}
\end{equation}
where $D_H(E)$ is the diffusion coefficient in the galactic halo.

As a matter of fact, radio observations in clusters of galaxies
yield for the corresponding diffusion constant $D_0$ 
a value similar to that found in the galactic
disk \cite{sst}
\footnote{Moreover, we note that average magnetic field 
values in galactic halos 
are expected to be close to those of galaxy clusters, 
i.e.  between 0.1 $\mu$G and 1 $\mu$G \cite{hillas}.}.
So, it looks plausible that a similar value for $D_0$ 
also holds on intermediate length scales, namely within the galactic halo.
In the lack of any further information on the energy-dependence of $D_H(E)$,
we assume the same dependence as that established for the disk.
Hence, from eq. (\ref{tau2})  
we find that for energies $E \ut < 10^3$ GeV the escape
time of CRs from the halo is greater than the age of the Galaxy
$t_0 \simeq 10^{10}$ yr
(notice that below the ultra-relativistic regime 
$\tau_{\rm esc}^{~H}$ gets even longer). As a consequence - since the CR flux 
scales like $E^{-2.7}$ (see next Section) - protons with $E \ut < 10^3$
GeV turn out to give the leading contribution to the CR flux.

We are now in position to evaluate the CR energy density in the galactic 
halo, getting
\begin{equation}
\rho_{CR}^{~H} \simeq \frac{3 t_0 L_G }{4 \pi R_H^3} \simeq 0.12
~~~~~{\rm eV~cm^{-3}}~,
\label{hcrd}
\end{equation}
where 
$L_G \simeq 10^{41}$ erg s$^{-1}$ 
is the galactic CR luminosity (see, e.g., \cite{breitschwerdt}).
Notice, for comparison, that $\rho_{CR}^{~H}$ turns out to be about
one-tenth of the disk value \cite{gaisser}.
In fact, this value is consistent with the EGRET upper bound on the CR 
density in the halo near the SMC \cite{sreekumar2}.

We remark that we have taken specific realistic values for the various
parameters entering the above equations in order to make a quantitative 
estimate.
However, somewhat different values can be used. For instance,
$R_H$ may range up to $\sim 200$ kpc \cite{bld},
whereas $D_0$ might be slightly larger than the above value, e.g. 
$\simeq 10^{29}$ cm$^2$ s$^{-1}$ 
consistently with our assumptions. Moreover, $L_G$ can be as large
as $3 \times 10^{41}$ erg s$^{-1}$ \cite{volk}. It is easy to see 
that these variations do not substantially affect our previous conclusions.

\section{Proton-proton scattering in the galactic halo}

We proceed to estimate the halo $\gamma$-ray flux produced by the
clouds clumped into dark clusters through the interaction
with high-energy CR protons. 
CR protons scatter on cloud protons giving rise (in particular) to neutral
pions, which subsequently decay into photons. 
A highly nontrivial question concerns the opacity effects in the clouds. 
Quite recently, Kalberla et al.~\cite{kalberla} have addressed precisely this 
issue, showing that optical-depth effects for both protons and photons are 
negligible within our model. 
Finally, we
expect an irrelevant high-energy ($\geq$ 100 MeV) $\gamma$-ray photon 
absorption outside the clouds, 
since the mean free path is orders of magnitudes larger than the 
halo size. 
 
As far as the energy-dependence of the halo CRs 
is concerned, we adopt the same power-law as in the galactic disk (see below)
\cite{gaisser}
\begin{equation}
\Phi^{H}_{CR}(E) \simeq \frac{A}{{\rm GeV}} 
\left(\frac{E}{{\rm GeV}}\right)^{-\alpha}~~~
{\rm particles~cm^{-2}~s^{-1}~sr^{-1}}~. \label{eqno:42}
\end{equation}
The constant $A$ is fixed by the requirement that the integrated
energy flux agrees with the above value of $\rho^H_{CR}$. Explicitly
\begin{equation}
\int d\Omega~ dE~ E~ \Phi^{H}_{CR}(E) \simeq 5.7\times 10^{-3}~~~
{\rm erg~cm^{-2}~s^{-1}}~,
\label{eqno:43}
\end{equation}
where for definiteness we take the integration range to be
$1~ {\rm GeV} \leq E \leq 10^3~ {\rm GeV}$.
A nontrivial point concerns the choice of $\alpha$. As an orientation, the observed
spectrum of primary CRs on Earth would yield 
$\alpha \simeq 2.7$. However, this conclusion cannot be extrapolated
to an arbitrary region in the halo (and in the disk), since 
$\alpha$ crucially depends on the diffusion processes undergone by
CRs. For instance, the best fit to EGRET data 
in the disk towards 
the galactic centre yields $\alpha \simeq 2.45$ \cite{mori},
thereby showing that $\alpha$ gets increased by diffusion.
In the lack of any direct information, 
we conservatively take $\alpha \simeq 2.7$ 
even in the halo,
but in Table 1 we report some results for different values 
of $\alpha$ for comparison . At any rate, the flux does not vary substantially.


\begin{table}
\caption{ 
Halo $\gamma$-ray intensity at high-galactic latitude 
for a spherical halo evaluated for $R_{min}= 10$ and 15 kpc
at energies above 0.1 GeV and 1 GeV, for different values of 
the CR spectral index $\alpha$ is given in units of
$10^{-7}$ $\gamma$  cm$^{-2}$ s$^{-1}$ sr$^{-1}$.} 
\begin{tabular}{cccc}
\br
$R_{min}   $ & $E_{\gamma}$ & $\alpha$  &  $\Phi_{\gamma}^{~\rm DM} 
(b=90\degr)$ \\
\hline
 (kpc)       &        (GeV) &           &  \\
\hline
\hline
$10$ & $>0.1$ &  2.45&   $62 $ \\
     &       &  2.70&   $59  $ \\
     &       &  3.00&   $49  $ \\
\hline
$10$ & $>1.0$  &  2.45&   $11$ \\
     &       &  2.70&   $6.7 $ \\
     &       &  3.00&   $3.3 $ \\
\hline
\hline
$15$ & $>0.1$ &  2.45&  $37 $ \\
     &       &  2.70&   $35 $ \\
     &       &  3.00&   $29 $ \\
\hline
$15$ & $>1.0$&  2.45&   $6.5$ \\
     &       &  2.70&   $4.0$ \\
     &       &  3.00&   $1.9$ \\
\br
\end{tabular}
\label{table3}
\end{table}

Let us next turn our attention to the evaluation of the $\gamma$-ray flux
produced in halo clouds
through the reactions $pp \rightarrow \pi^0 \rightarrow \gamma
\gamma$. Accordingly, the source function 
$q_{\gamma}(>E_{\gamma},\rho,l,b)$ -
yielding the photon number density at distance
$\rho$ from Earth with energy greater than $E_{\gamma}$ - is 
\cite{gaisser}
\begin{equation}
\begin{array}{ll}
q_{\gamma}(>E_{\gamma},\rho,l,b)= 
\displaystyle{\frac{4\pi}{m_p}}  \rho_{H_2}(\rho,l,b)~\times \\ \\
\sum_{n} \int_{E_p(E_{\gamma})}^{\infty}
d{E}_p  dE_{\pi}  \Phi^H_{CR}
({E}_p) 
\displaystyle{\frac{d \sigma^n_{p \rightarrow \pi}
(E_{\pi})}{dE_{\pi}}}
n_{\gamma}({E}_p)~, 
\label{eqno:46}
\end{array}
\end{equation}
where 
the lower integration limit $E_p(E_{\gamma})$ is the minimal proton 
energy necessary to produce a photon with energy $>E_{\gamma}$,
$\sigma^n_{p \rightarrow \pi}(E_\pi)$ is the cross-section for the
reaction $pp \rightarrow n \pi^0$ ($n$ is the $\pi^0$ multiplicity), 
$\rho_{H_2}(\rho,l,b)$ is the halo gas density profile
and $n_{\gamma}({E}_p)$ is 
the photon multiplicity.

Unfortunately, it would be exceedingly difficult to keep track of the
clumpiness of the actual gas distribution in the halo, and 
so we assume that its
density is smooth and goes like the dark matter density - anyhow, the very low 
angular resolution of $\gamma$-ray detectors would not permit to
distinguish between the two situations (evidently this strategy would be 
meaningless if optical-depth effects were not negligible). 
Accordingly, the halo gas
density profile reads  
\begin{equation}
\rho_{H_2}(x,y,z) = f~ {\rho_0 (q)} ~ \frac{\tilde a^2+R_0^2}{\tilde a^2+x^2+y^2+(z/q)^2}~,
\label{eqno:29}
\end{equation}
for $\sqrt{ x^2+y^2+z^2/q^2} > R_{min}$, 
($R_{min} \simeq 10$ kpc is the minimal galactocentric distance of the dark clusters 
in the galactic halo). 
We recall that $f$ denotes the fraction of halo dark matter in the form of gas,
$\rho_0(q)$ is the local dark matter density,
$\tilde a = 5.6$ kpc is the core radius and $q$ measures 
the halo flattening. For 
the standard spherical halo model 
$\rho_0(q=1) \simeq 0.3$ GeV cm$^{-3}$,
whereas it turns out that e.g. 
$\rho_0(q=0.5) \simeq 0.6$ GeV cm$^{-3}$.

In order to proceed further, it is convenient
to re-express 
$q_{\gamma}(>E_{\gamma},\rho,l,b)$ in terms of the inelastic pion production
cross-section $\sigma_{in}(p_{lab})$. Since
\begin{equation}
\sigma_{in}(p_{lab})<n_{\gamma}(E_p)>~ = \sum_n \int dE_{\pi}~
\frac{d\sigma^n_{p \rightarrow \pi}(E_{\pi})}{dE_{\pi}}~ n_{\gamma}(E_p)~,
\label{eqno:48}
\end{equation}
eq. (\ref{eqno:46}) becomes
\begin{equation}
\begin{array}{ll}
q_{\gamma}(>E_{\gamma},\rho,l,b)= 
\displaystyle{\frac{4\pi}{m_p}}
\rho_{H_2}(\rho,l,b) ~\times \\ \\
\int_{E_p(E_{\gamma})}^{\infty}
d{E}_p~ \Phi^H_{CR}({E}_p)~ 
\sigma_{in}(p_{lab}) <n_{\gamma}({E}_p)>~, \label{eqno:49}
\end{array}
\end{equation}
where $\rho_{H_2}(\rho,l,b)$ is given by eq. (\ref{eqno:29}) with 
$x = -\rho \cos b \cos l +R_0$,
$y = -\rho \cos b \sin l$ and 
$z =  \rho \sin b$.

For the inclusive cross-section of the reaction 
$pp \rightarrow  \pi^{0}  \rightarrow \gamma \gamma$ 
we adopt the Dermer \cite{dermer} parameterization 
\begin{equation}
\begin{array}{ll}
\sigma_{in}(p) < n_{\gamma}(E_p) > = 2 \times 1.45 \times
10^{-27} ~\times \\ \\
~~~\left\lbrace 
\begin{array}{llll} 
0.032 \eta^2 + 0.040 \eta^6 + 0.047 \eta^8 & ~~~0.78 \leq p \leq 0.96 \\
32.6 (p - 0.8)^{3.21}                      & ~~~0.96 \leq p \leq 1.27 \\ 
5.40 (p - 0.8)^{0.81}                      & ~~~1.27 \leq p \leq 8.0 \\
32 ln p + 48.5 p^{-1/2} - 59.5             & ~~~p \geq 8.0 ~,
\end{array}
\right. 
\end{array}
\end{equation}
where $p$ is the proton laboratory momentum in GeV/c,
the factor 2 comes from the fact that each pion decays into two photons,
whereas 1.45 accounts for the CR composition \cite{dermer},
which includes also heavy nuclei.
The quantity 
\begin{equation}
\eta \equiv \frac{[(s- m_{\pi}^2 - m_p^2)^2 - 4m_{\pi}^2  m_p^2]^{1/2}}
{2 m_{\pi} s^{1/2} }~,
\end{equation}
is expressed in terms of the Mandelstam
variable $s$, while $m_{\pi}$ and $m_p$ are the pion
and the proton mass, respectively. 

Because $dV=\rho^2 d\rho d\Omega$, it follows that the observed
$\gamma$-ray flux per unit solid angle is
\begin{equation}
\Phi_{\gamma}^{~ \rm DM}
(>E_{\gamma},l,b)=\frac{1}{4\pi} 
\int^{\rho_2(l,b)}_{\rho_1(l,b)} d\rho~ q_{\gamma}(>E_{\gamma},\rho,l,b)
~. \label{eqno:51}
\end{equation}
So, we find 
\begin{equation}
\Phi_{\gamma}^{~ \rm DM}
(>E_{\gamma},l,b) = 
f ~ \frac{\rho_0(q)}{m_p}~ {I}_1(l,b)~ 
{I}_2(>E_{\gamma})~, 
\label{eqno:52}
\end{equation}
where
${I}_1(l,b)$ and ${I}_2(>E_{\gamma})$ are
defined as
\begin{equation}
{I}_1(l,b) \equiv \int^{\rho_2(l,b)}_{\rho_1(l,b)} d\rho 
\left(\frac{\tilde a^2 + R_0^2}
{\tilde a^2 + x^2 + y^2 + (z/q)^2 } \right)~, \label{eqno:A5}
\label{eqno:39}
\end{equation}
\begin{equation}
{I}_2(>E_{\gamma}) \equiv \int_{E_p(E_{\gamma})}^{\infty} 
d{E}_p~\Phi^H_{CR}({E}_p)~\sigma_{in}(p_{lab})
<n_{\gamma}({E}_p)>~, 
\label{eqno:35}
\end{equation}
and $m_p$ is the proton mass.

According to the discussion in Sections 2 and 3, 
typical values of $\rho_1(l,b)$ and $\rho_2(l,b)$ in eqs. 
(\ref{eqno:51}) and (\ref{eqno:39})
are 10 kpc and 100 kpc, respectively.
Numerical values for $\Phi_{\gamma}^{~\rm DM}$ in the 
cases $\alpha = 2.45,~ 2.7$ and $3.0$ are reported in Table 1.

\section{Inverse-Compton scattering}

Another mechanism  whereby $\gamma$-ray photons are produced is 
IC scattering of high-energy CR electrons off galactic background photons. 
Here we estimate the resulting flux, while the interplay between 
proton-proton scattering and IC scattering will be discussed in the next Section.


The electron injection spectrum which best fits the locally observed 
electron spectrum is given by the following power-law 
valid for $E_e \ut > 10$ GeV
(see e.g. \cite{porter})
\begin{equation}
I_e(E_e;\rho,l,b) = K(\rho,l,b) E_e^{-a}~~
{\rm e^-~cm^{-2}~s^{-1}~sr^{-1}~Gev^{-1}}~,
\label{11}
\end{equation}
with $a \simeq 2.4$ and 
$K_0 \equiv K(0) \simeq 6.3 \times 10^{-3}$ 
e$^-$ cm$^{-2}$ s$^{-1}$ sr$^{-1}$ Gev$^{a-1}$ (the value of $K_0$ is
obtained by normalizing eq. (\ref{11}) with the observed local CR
electron spectrum at 10 GeV).
Since $a$ is somewhat model-dependent (in particular it depends
on the diffusion 
processes), its actual value is not well determined, and indeed it could 
be as low as $a \simeq 2$ \cite{pohl} or even $a\simeq 1.8$ \cite{smr}.
However, what is relevant is the electron spectrum where
the $\gamma$-ray production occurs and - due to diffusion processes -
the value of $a$ is expected to increase with the distance from the 
galactic plane where the electrons are mostly produced.

In order to estimate the 
galactic radiation field, we adopt the model of Mazzei,
Xu \&  De Zotti \cite{mazzei}
for the photometric evolution of disk galaxies. This model reproduces well 
the present broad-band spectrum of the Galaxy over about four decades 
in frequency, from UV to far-IR. Accordingly, the two main 
contributions to the galactic radiation field come from stars at wavelength 
$\lambda$ $\sim 1 \mu$m and diffuse dust at $\lambda \sim 100 \mu$m.
The total stellar luminosity of the Galaxy is  
$L_\star \sim 3.5 \times 10^{10}~L_{\odot}$ 
and the amount of starlight absorbed and re-emitted by dust is 
$L_{\rm d} \sim 1.2 \times 10^{10}~L_{\odot}$
(see e.g., \cite{mazzei,cox}).
As regards to the photon energy distribution, we can roughly approximate
the emission spectrum (see Fig. 4 in \cite{mazzei})
with the sum of two Planck functions
with temperature $T_{\star} \sim 2900$ K and $T_{\rm d} \sim 29$ K,
respectively.

According to the previous assumptions,
the source function 
$q_{\rm ph} (E_{\gamma})$ for $\gamma$-ray
production through IC scattering is given by
\cite{berezinskii} 
\begin{equation}
\begin{array}{ll}
q_{\rm ph} (E_{\gamma}) =
\frac{1}{2} \sigma_T  
\left( \frac{4}{3} <\epsilon_{\rm ph}(T_{\star,{\rm d}})> \right)^{(a-1)/2} 
~\times \\ \\
(mc^2)^{1-a} K_0 E_{\gamma}^{-(a+1)/2} 
~~~{\rm \gamma~ s^{-1}~sr^{-1}~Gev^{-1}}.
\label{la}
\end{array}
\end{equation}
Here $<\epsilon_{\rm ph}(T_{\star,{\rm d}})> 
\simeq 8kT_{\star,{\rm d}}/3$
is the average energy of background photons emitted by stars or dust and 
$\sigma_T$ is the Thompson cross-section. In deriving eq. (\ref{la}), use 
is made of the fact that the $\gamma$-ray energy is related to 
the electron and background photon energies
according to
\begin{equation}
<E_{\gamma}> = \frac{4}{3} <\epsilon_{\rm ph}(T_{\star, d})> \left( \frac{ E_e}{mc^2} \right)
^2~,
\label{egammaav}
\end{equation}
so that very high-energy electrons are needed in order to produce 
$\gamma$-rays. 
For example, a $\gamma$-ray with $E_{\gamma} \simeq 1$ GeV produced by this 
mechanism requires $E_e \simeq 170$ GeV for a target photon emitted by dust, 
while $E_e \simeq 17$ GeV is demanded for starlight.

The intensity of diffuse galactic $\gamma$-rays of energy 
$>E_{\gamma}$ produced in this way and coming to Earth along 
the line-of-sight
$(l,b)$ turns out to be
\begin{equation}
\begin{array}{ll}
\Phi_{\gamma}^{~\rm IC} (>E_{\gamma},l,b)  = 
\int_{0}^{\infty} d\rho <n_{\rm ph}~(\rho,l,b)>f_e(\rho,l,b) ~ \times \\ \\
\int_{E_{\gamma}}^{\infty} 
q_{\rm ph }({E}_{\gamma})~ d{E}_{\gamma}
{\rm ~~~\gamma~cm^{-2}~s^{-1}~sr^{-1}}~,
\label{eqno:intensityph}
\end{array}
\end{equation}
where we have introduced the function
$f_{e}(\rho,l,b) \equiv K(\rho,l,b)/K_0$ 
as the ratio of the electron CR intensity 
relative to the local intensity, while
$<n_{\rm ph}(\rho,l,b)>$ is
the average density of background photons.

Let us next focus our attention on 
the functions $f_{e}(\rho,l,b)$ and $<n_{\rm ph}(\rho,l,b)>$.
The electron component of CRs is galactic in 
origin, mainly produced by supernovae and pulsars located 
inside the disk.
Electrons diffuse through the Galaxy and their distribution is 
energy-dependent and not uniform, namely,
the characteristic diffusion length scale gets smaller
for higher electron energy. This feature  
cannot be described in the framework 
of the widely used Leaky Box Model, and in order to obtain the electron 
density at an arbitrary point in the Galaxy 
one has to resort to the transport equation (see e.g.  
\cite{pohl,berezinskii}).
Unfortunately, several fairly unknown parameters enter this equation,
like the electron diffusion coefficient, 
the rate at which electrons lose energy, the density of sources and the 
electron spectrum.

An alternative approach relies upon the experimental evidence of the thick
disk
\footnote{Often defined as ``halo'' by the CR community.},  
in which high-energy electrons 
may be retained for a long time before escaping into the galactic halo.
Indeed, the observed characteristics of the radio emission 
spectra of our and other galaxies lead to a relative density distribution of 
electrons $f_e(R_0,z)\equiv n_e(z)/n_e(0)$ extending  up to $5 -12$ kpc 
perpendicularly to the galactic plane, as shown 
in Figure 5.29 of \cite{berezinskii}. 
These numerical results can be 
approximated by $f_e(R_0,z) = \exp[-(z/z_e)^{3/2}]$, 
with the parameter $z_e$ depending on the electron energy. From 
eq.~(\ref{egammaav}) and the ensuing discussion, it turns out that
$z_e \simeq 2.5$ kpc for $E_e \simeq 170$ GeV
while
$z_e \simeq 3.5$ kpc for $E_e \simeq 17$ GeV. 
As far as the radial dependence 
of the electron distribution is concerned, 
we assume that $f_{e}(R,0)$ follows the same 
$R$-dependance of the CRs, which can be obtained by 
using a best fit procedure to the data in  
Figure 11 of \cite{bloemen}. This yields
\begin{equation}
f_{e}(R,0) = e^{[0.48-0.36(R/R_0)-0.12(R/R_0)^2]}~.
\label{eqno:4}
\end{equation}   
However, following Bloemen \cite{bloemen} - who 
suggested a stronger radial gradient for the electron component of the CRs
- we also tested the effect of using a 
steeper radial electron distribution on the IC $\gamma$-ray flux.  
We anticipate that the corresponding results show that the IC 
$\gamma$-ray flux does not change
significantly for galactic longitudes $|l| \leq 90^0$ (irrespectively of
the latitude values) while it increases up to a factor of two at
$l = 180^0$ for $|b| \leq 30^0$.

The last quantity to be specified 
in eq. (\ref{eqno:intensityph}) is the average 
background photon density $<n_{ph}(\rho, l, b)>$ or, equivalently, 
the background photon flux $\Phi_{ph}(\rho, l, b)$ emitted by stars and dust
\begin{equation}
<n_{\rm ph}(\rho,l,b)>~ = \frac{\Phi_{\rm ph}(\rho,l,b)}{c} ~~~
{\rm \gamma~cm^{-3}}~.
\label{eqno:16}
\end{equation}
Note that the photon flux $d \Phi_{\rm ph}(\rho,l,b)$ at a point
$P(\rho, l, b)$  from 
the solid angle $d\Omega$ 
subtended by an infinitesimal area $d{a'}$ centered in 
$P'(R',\phi', z'=0)$ on the galactic plane is given by
\begin{equation}
d\Phi_{\rm ph}(\rho,l,b) =
I_{*,d}(R')~\left(\frac{d\Omega}{4\pi}\right)\cos{\alpha}~~~
{\rm ~~~\gamma~cm^{-2}~s^{-1}}~,
\label{eqno:17}
\end{equation}
where $\alpha$ is the angle between the normal to the area $da'$ and
the direction  ${\bf PP'}$.
We can trace the surface brightness $I_{\star,{\rm d}}(R')$ to
the stellar/dust distribution. Assuming that visible matter makes up an 
exponential disk, we set
\begin{equation}
I_{*,d} (R') = A_{*,d} e^{-(R'-R_0)/h_{*,d}}~~~~~~
{\rm \gamma ~ cm^{-2}~ s^{-1}} ~,
\label{22}
\end{equation}
where $h_{*,d}\simeq 3.5$ kpc is the scale length for the visible matter 
and the constant $A_{*,d}$ is fixed by the total disk luminosity as
\begin{equation}
\int_0^{R_d} I_{*,d}(R') 2 \pi R' dR' = 
\frac {L_{\star,d}} {2<\epsilon_{\rm ph}(T_{\star,d})>} 
~~~~  {\rm \gamma ~s^{-1}}~.
\end{equation}
In this way, we get  
$A_{\star} = 4.71 \times 10^{20}$ $\gamma$ cm$^{-2}$ s$^{-1}$
and 
$A_{d}     = 1.64 \times 10^{22}$ $\gamma$ cm$^{-2}$ s$^{-1}$,
with $R_d \simeq 15$ kpc.
By integrating eq. (\ref{eqno:17}) on the galactic disk, we find
\begin{equation}
\Phi_{\rm ph}(\rho,l,b)= 
\int_0^{R_d} 
\int_0^{2\pi} 
I_{*, d}(R')~R' dR' d\phi '~
\left(\displaystyle{\frac{\cos\alpha}{4 \pi |{\bf PP'}|^2}}\right) ~~~~
{\rm \gamma~cm^{-2}~s^{-1}} ~.
\label{24}
\end{equation}
Finally, by using eqs. (\ref{eqno:16}), (\ref{22}) and (\ref{24}) - and 
recalling eq. (\ref{la}) - eq. (\ref{eqno:intensityph}) can be rewritten in the form
\begin{equation}
\Phi_{\gamma}^{~\rm IC}(>E_{\gamma},l,b)  = 
{J}_1(l,b)~{J}_2(>E_{\gamma}) 
~~~~{\rm \gamma~cm^{-2}~s^{-1}~sr^{-1}}~,
\end{equation}
where we have set
\begin{equation}
\begin{array}{ll}
{J}_1(l,b) \equiv
\int_{0}^{\infty} f_e(\rho,l,b) d\rho ~\times \\ \\
\int_0^{R_d}~\int_0^{2 \pi} 
\left(\displaystyle{
\frac{\cos\alpha} {4 \pi |{\bf PP'}|^2}} \right) ~R' dR' d\phi ' 
e^{-(R'-R_0)/h_{*,d}}~~~{\rm ~cm}~,
\end{array}
\end{equation}
and
\begin{equation}
\begin{array}{ll}
{J}_2(>E_{\gamma}) \equiv 
\displaystyle{\frac{A_{*,d}}{2c}}
 \sigma_T~ [ 4/3 <\epsilon_{\rm ph}(T_{\star,d})>]^{(a-1)/2}~\times
\\ \\
~(mc^2)^{1-a}~K_0~
\int_{E_{\gamma}}^{\infty} {E}_{\gamma}^{-(a+1)/2}d{E}_{\gamma}
~~~~~~~{\rm \gamma~cm^{-3}~s^{-1}~sr^{-1} }~.
\end{array}
\label{eqno:25}
\end{equation}

Numerical values of $\Phi_{\gamma}^{~\rm IC}(>E_{\gamma},l,b)$ at 
high-galactic latitude are exhibited in Table 2 and plotted in Figure 3.
   \begin{table}
      \caption{ 
The galactic diffuse $\gamma$-ray intensity due to IC scattering 
of high-energy electrons on background photons from stars and dust
is given (in units of $10^{-7}$ $\gamma$ cm$^{-2}$ s$^{-1}$ sr$^{-1}$)
for $a=2.0,~2.4$ and $2.8$. The results for $a=2, ~2.8$ are reported
for illustrative purposes.
We adopt the following values: $T_* = 2900$ K, 
$L_* = 3.5 \times 10^{10}~L_{\odot}$ and
$T_d = 29$ K, $L_d= 1.5 \times 10^{10}~L_{\odot}$.}
\begin{tabular}{cccccc}
\br
 & $z_e$ & $E_{\gamma}$ &
$\Phi_{\gamma}^{~\rm IC}(90^0)$&
$\Phi_{\gamma}^{~\rm IC}(90^0)$&
$\Phi_{\gamma}^{~\rm IC}(90^0)$ \\ 
\hline
  & (kpc) & (GeV) & & & \\
\hline
\hline
 & & & $a=2.0$ & $a=2.4$ & $a=2.8$ \\
\hline
stars  & 3.5& $>0.1$ &$3.8 $ &  $3.5 $&  $3.4 $ \\
\hline
       &    & $>1.0$ &$1.2 $ &  $0.7 $&  $0.4$ \\
\hline
\hline
dust   & 2.5& $>0.1$ &$12$ &  $4.4 $&  $1.7 $ \\
\hline
       &    & $>1.0$ &$3.8 $ &  $0.9$&  $0.2$ \\
\br
\end{tabular}
\label{table2}
\end{table}

\section{Discussion}
\begin{figure*} 
\vspace{15.cm}
\includegraphics{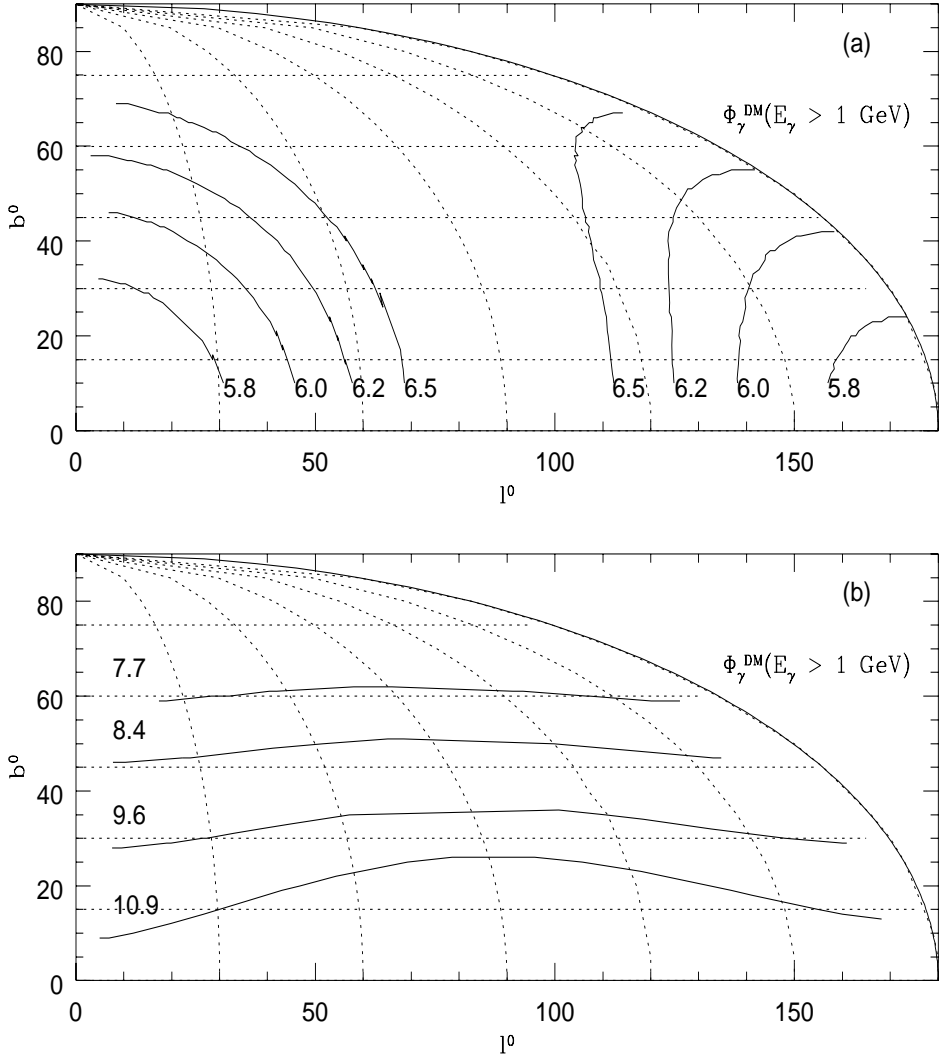} 
\caption{Contour values for the $\gamma$-ray flux due 
to the DM at $E_{\gamma} > 1$ GeV are given for the indicated values 
in units of $10^{-7}$ $~\gamma$ cm$^{-2}$ s$^{-1}$ sr$^{-1}$,
in the cases: (a) spherical halo, (b) flattened halo with $q=0.5$.}
\label{fig2}
\end{figure*}
Our main result are maps for the intensity distribution 
of the $\gamma$-ray emission from 
baryonic dark matter (DM) in the galactic halo and from IC 
processes in the galactic disk. In order to make the discussion definite,
we take the fraction of halo dark matter in the form of molecular clouds
$f \simeq 0.5$. As far as the IC emission is concerned, the 
standard electron spectral index $a=2.4$ is used.
We stress that the shape of the IC maps does not depend on the value of $a$.



In Figures 1 we exhibit the contour plots in the first quadrant of the 
sky ($0\degr \le l \le 180\degr$, $0\degr \le b \le 90\degr$) for 
the halo $\gamma$-ray flux 
$\Phi_{\gamma}^{~\rm DM} (E_{\gamma}> 1 
{~\rm GeV})$.
Corresponding contour plots for $E_{\gamma}>0.1$ GeV are identical,
up to an overall constant factor equal to 8.74, as follows from
eq. (\ref{eqno:52}).

Figure 1a refers to a spherical halo, whereas Figure 1b 
pertains to a $q=0.5$ flattened halo.
Regardless of the adopted value for $q$,
$\Phi_{\gamma}^{~\rm DM}(E_{\gamma}>1{\rm ~GeV})$ lies in the range
$\simeq 6-8 \times 10^{-7}$ $\gamma$ cm$^{-2}$ s$^{-1}$ sr$^{-1}$
at high-galactic latitude. 
However, the shape of the contour lines strongly depends on
the flatness parameter. 
Indeed, for $q \ut > 0.9$ there are two contour lines (for 
each flux value) approximately symmetric with respect to $l=90\degr$
(see Figure 1a). 
On the other hand, for 
$q \ut < 0.9$ there is a single contour line (for each value of the flux) 
which varies much less with the longitude (see Figure 1b). 

As we can see from Table 1 and Figures 1, the predicted
value for the halo $\gamma$-ray flux at high-galactic latitude
is close to that found by Dixon et al. \cite{dixon} (see also Table 3).
This conclusion holds almost 
irrespectively of the flatness parameter. 

Moreover, the comparison of the overall shape of the contour lines in our 
Figures 1a and 1b with the corresponding ones in Figure 3 of Ref.
\cite{dixon} entails that models 
with flatness parameter
$q \ut < 0.8$ are in better agreement with the data, 
thereby implying that most likely 
the halo dark matter is not spherically distributed. This result has  
been also recently confirmed in the analysis  by \cite{kalberla}.

In Figure 2 we present contour plots for the $\gamma$-ray flux
due to the IC scattering, for $E_{\gamma}>1$ GeV. 
The corresponding contour plots for $E_{\gamma}>0.1$ GeV are identical,
up to an overall constant factor equal to 5
(this follows from eq. (\ref{eqno:25})).
The contour lines decrease with increasing longitude.
\begin{figure*} 
\vspace{8.cm}
\includegraphics{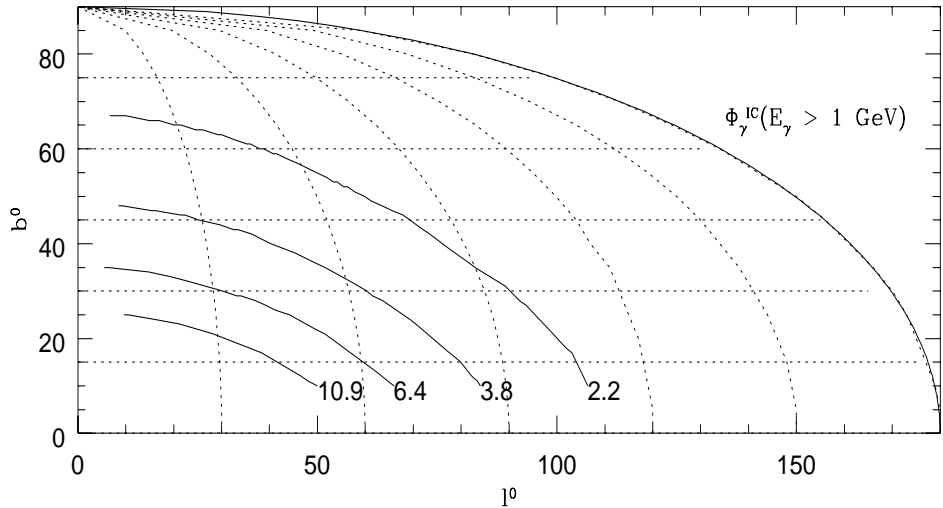} 
\caption{Contour values for the $\gamma$-ray flux due to the IC 
at $E_{\gamma} > 1$ GeV are given for the indicated values 
in units of $10^{-7}$ $~\gamma$ cm$^{-2}$ s$^{-1}$ sr$^{-1}$.}
\label{fig3}
\end{figure*}

We remark that eq. (\ref{eqno:52}) yields 
$\Phi_{\gamma}^{~DM}(E_{\gamma}>0.1{~\rm GeV}) \simeq 5.9 \times 10^{-6}$
$~\gamma$ cm$^{-2}$ s$^{-1}$ sr$^{-1}$
at high-galactic latitude (for a spherical halo). This value is roughly
40\% of the diffuse $\gamma$-ray emission of 
$(1.45 \pm 0.05)  \times
10^{-5}~\gamma~{\rm cm^{-2}~s^{-1}~sr^{-1}}$
found by the EGRET team \cite{sreekumar} and in agreement with
the conclusion of Dixon et al. \cite{dixon} that 
the halo $\gamma$-ray emission is a relevant fraction of the 
standard isotropic diffuse flux also for $E_{\gamma} > 0.1$ GeV.
   \begin{table}
      \caption{
Rough values of the measured residual $\gamma$-ray flux at 
$E_{\gamma}\geq 1$ GeV (after subtraction of both the isotropic background 
and the standard galactic diffuse component) 
is given for different galactic latitude and longitude values 
(interpolated from Fig. 3a in [1]).
Fluxes are given in units of $10^{-6}$ $\gamma$ cm$^{-2}$ s$^{-1}$ sr$^{-1}$.
} 
\begin{tabular}{ccc}
\br
$b$ & $l=0\degr$ & $l=60\degr$  \\
\hline
\hline
$45\degr$ & 1 &  1 \\
\hline
$30\degr$ & 2 &  1.5 \\
\hline
$15\degr$ & 5.5 &  2 \\
\br
\end{tabular}
\label{table_obs}
\end{table}

Nevertheless, given the large 
uncertainties both in the data and in the model parameters
(such as for instance the electron scale height and  
the electron spectral index $a$), one might also explain the 
observations with a nonstandard IC 
mechanism \cite{smr}. 
Our calculation, however, seems to point out that the IC
contour lines in Figure 2
decrease much more rapidly than the observed ones for the halo
$\gamma$-ray emission (see Figure 3 in \cite{dixon}). 
More precise measurements with a next generation
of satellites are certainly required in order to settle the issue.

\section{Gamma rays from the halo of M31}
As M31 resembles our galaxy, the discovery of Dixon et al. \cite{dixon} 
naturally leads to the expectation that the halo of M31 should give rise to 
a $\gamma$-ray emission as well. Below, we will try to address this issue 
in a quantitative manner, assuming that the halo of M31 is structurally 
similar to that of our galaxy and that our model for baryonic 
dark matter is correct.

We suppose that the various parameters entering the calculations in 
Sections 3 and 4 take similar values for M31 and for the Galaxy, apart 
from the M31 central dark matter density 
$\rho(0) \simeq 2.5 \times 10^{-24}$ g cm$^{-3}$
and the M31 core radius $\tilde a \simeq 5$ kpc.
Accordingly, the evaluation of the corresponding flux 
$\Phi^{~M31}_{\gamma~~halo}$ proceeds as before,
with only minor modifications.
Specifically, we can use again eq. (\ref{eqno:52}) - with 
${I}_2$ still given by eq. (\ref{eqno:35}) - but now 
${I}_1$ is to be replaced by
${L}_1$ (see below), in order to account for the different geometry.
Notice that $f$ in eq. (\ref{eqno:52}) presently denotes the fraction of 
halo dark matter of M31 in the form of $H_2$ clouds.

Consider a generic point $P$ in the halo of M31, and let 
$R$ and $r$ denote its distance from the centre $O$ of M31 and from Earth, 
respectively.
Since the distance of $O$ from Earth is $D \simeq 650$ kpc, we have
$R(r) = (r^2 + D^2 - 2 r D cos \theta)^{1/2}$, where $\theta$ 
is the angular separation between $P$ and $O$ as seen from Earth.
For simplicity, we suppose that the M31 halo is described by an isothermal 
sphere with radius $R_H$ and density profile
\begin{equation}
\rho(R) = \frac{\rho(0)}{1+(R/\tilde a)^2}~.
\end{equation}
Note that the ensuing amount of dark matter in M31 turns out to be about twice 
as large as that of the Galaxy.

According to the discussion in Section 2, the dark clusters should populate 
only the outer halo of M31. So, we compute $\Phi^{~M31}_{\gamma~~halo}$ from 
regions of the M31 halo with 
$R_{min} < R < R_H$, with $R_{min} \simeq 10$ kpc and $R_H \simeq 100$ kpc, 
for definiteness. As it is easy to see, the values of $\theta$ 
corresponding to $R_{min}$ and $R_H$ are $\theta_{min} \simeq 1\degr$ and
$\theta_{H} \simeq 9\degr$, respectively.

We are now in position to compute ${L}_1$, which reads
\begin{equation}
{L}_1 = 2\pi
\int_{\theta_{min}}^{\theta_{H}}
\sin \theta ~d\theta
\int_{r_{min}(\theta)}^{r_{max}(\theta)} 
dr 
\left( \displaystyle{\frac{\tilde a^2}{\tilde a^2 +
R^2(r)}} \right ) \simeq 
~1.9 \times 10^{20}~ {\rm cm~sr}~,
\label{eqno:m3144}
\end{equation}
with $r_{max (min)}(\theta) \equiv D cos \theta + (-) 
(R^2_H - D^2 sin^2 \theta)^{1/2}$.
Recalling eqs. (\ref{eqno:52}) and (\ref{eqno:35}),
we get 
\begin{equation}
\Phi^{~M31}_{\gamma~~halo}(>E_{\gamma}) = 
1.9 \times 10^{20} f \frac{\rho(0)}{m_p} 
{I}_2(>E_{\gamma}) 
~~~{\rm \gamma~ cm^{-2}~s^{-1}}~. 
\label{eqno:m3145}
\end{equation}
Observe that regions of M31 halo with angular separation less than
$\theta_{min}$ from $O$ do not contribute in eqs. (\ref{eqno:m3144}) 
and (\ref{eqno:m3145}), and so $\Phi^{~M31}_{\gamma~~halo}$
should be regarded as a lower bound on the total $\gamma$-ray flux from M31 
halo.

Specifically, eq. (\ref{eqno:m3145})
yields 
\begin{equation}
\Phi^{~M31}_{\gamma~~halo}(E_{\gamma}>0.1{~\rm GeV}) 
\simeq 3.5 \times 10^{-7} f 
~~~{\rm \gamma~ cm^{-2}~s^{-1}}~.
\label{m31} 
\end{equation}

This value has to be compared both with the $\gamma$-ray flux
from M31 disk 
and with the $\gamma$-ray emission from the halo of the Galaxy.
The former quantity has been estimated to be $\simeq 0.2 \times 10^{-7}$ 
$\gamma$  cm$^{-2}$ s$^{-1}$ for $E_{\gamma}>0.1$ GeV \cite{ob,of} 
within a field of view of $1.5\degr\times 6\degr$, 
whereas the latter quantity, integrated over the entire field of view of 
M31 halo, is $\simeq 4.3 \times 10^{-7}$ 
$\gamma$  cm$^{-2}$ s$^{-1}$ for $E_{\gamma}>0.1$ GeV, according to our 
results in  Section 4 and 6. \footnote{For simplicity, we suppose here that 
the halo of the Galaxy is spherical and we employ eq. (\ref{eqno:52}) 
with $f=1/2$.}

As far as observation is concerned, no $\gamma$-ray flux from M31 has been 
detected by EGRET. Accordingly, the EGRET team has derived the upper bound
\cite{sreekumar94}
\begin{equation}
\Phi^{~M31}_{\gamma}(E_{\gamma}>0.1{~\rm GeV}) \ut <0.8 \times 10^{-7}  
~~~{\rm \gamma~ cm^{-2}~s^{-1}}~.
\label{m31obs}
\end{equation}

Unfortunately, a direct comparison between eqs. (\ref{m31}) and
(\ref{m31obs}) is hindered by the fact that eq. (\ref{m31obs}) is derived 
under the assumption of a point-like source.


Clearly, a good angular resolution of about one degree or less is  
necessary in order to discriminate between the halo and disk emission from M31. 
So, the next generation of $\gamma$-ray satellites 
like AGILE and  GLAST can test 
our predictions.

\ack{We would like to thank G. Bignami, P. Caraveo, D. Dixon,
T. Gaisser, M. Gibilisco,
G. Kanbach, T. Stanev, A. Strong and M. Tavani for useful discussions.}

\end{document}